\documentclass[twocolumn,pra,aps,showpacs]{revtex4}
\usepackage{epsfig}

\def\beq{\begin{equation}}
\def\eeq{\end{equation}}
\def\bea{\begin{eqnarray}}
\def\eea{\end{eqnarray}}

\newcommand{\la}[1]{\label{#1}}
\newcommand{\ur}[1]{(\ref{#1})}
\newcommand{\urs}[2]{(\ref{#1},\ref{#2})}
\newcommand{\eq}[1]{Eq.~(\ref{#1})}

\newcommand{\half}{\frac{1}{2}}

\begin{document}
\title{Loop structure of the lowest Bloch band for a Bose-Einstein condensate}
\author{Dmitri Diakonov$^{1,2}$, L.\ M.\ Jensen$^1$,
C.\ J.\ Pethick$^1$, and H.\ Smith$^3$}

\affiliation{$^1$ NORDITA, Blegdamsvej 17, DK-2100 Copenhagen \O,
Denmark\\
$^2$ St. Petersburg Nuclear Physics Institute,
Gatchina, St. Petersburg 188300, Russia \\
$^3$ {\O}rsted Laboratory,
 H.\ C.\ {\O}rsted Institute,
Universitetsparken 5, DK-2100 Copenhagen {\O},
Denmark}

\date{\today}

\begin{abstract}

We investigate analytically and numerically Bloch waves for a Bose--Einstein
condensate in a sinusoidal external potential.  At low densities the
dependence of the energy on the quasimomentum is similar to that for a single
particle, but at densities greater than a critical one the lowest band
becomes triple-valued near the boundary of the first Brillouin zone and
develops the structure characteristic of the swallow-tail catastrophe.
We comment on the experimental consequences of this behavior.
\end{abstract}



\maketitle

The possibility of studying experimentally the properties of Bose-Einstein
condensates in periodic potentials created by optical lattices (see e.\ g.\ 
\cite{expt, morsch}) has stimulated a number of theoretical
investigations.  These include nonlinear tunneling phenomena \cite{WN1, ZG},
Bloch oscillations \cite{kirstine, choi}, stability of Bloch waves \cite{WN2},
and oscillations of condensates and localized excitations \cite{TS}.  In a
two-level model of nonlinear Landau-Zener tunneling, evidence was found for a
loop structure in the adiabatic energy levels \cite{WN1}.  Recently Bronski 
{\it
et al.}  \cite{BCDK} discovered a remarkable exact solution of the
Gross-Pitaevskii equation for a condensate in a one-dimensional optical
lattice for a quasimomentum corresponding to the boundary of the first
Brillouin zone.  The reason for an exact solution being possible is that the
spatially dependent parts of the lattice potential and the internal field due
to interparticle interactions cancel.  Subsequently, in Ref.\ \cite{WDN} it
was shown by numerical studies that the current of the condensate for
quasimomenta in the vicinity of the zone boundary had a singular behavior, and
it was speculated that the chemical potential of the condensate became
multivalued near the zone boundary.  In this Letter we use analytical and
numerical methods to show that the lowest band has the structure associated
with the swallow-tail catastrophe \cite{thom}, and we comment on implications
of this result for experiment.

The potential $V(x)$ produced by a one-dimensional standing light wave has the 
form
\beq
V(x)=V_0\,\sin^2 \frac{\pi x}{d}, \la{V}
\eeq
where $d$ is the period of the lattice.
We shall assume that the condensate
may be described by the Gross--Pitaevskii functional for the energy density,
which when averaged over a lattice period is given by
\beq
{\cal E}=\frac{1}{d}\int_{-d/2}^{d/2}
dx\left[\frac{\hbar^2}{2m}\left|\frac{d\psi}{dx}\right|^2
+V(x)|\psi|^2+\frac{U_0}{2}|\psi|^4\right],
\la{calE}\eeq
and the mean number density $n$ is given by
\beq n=\frac{1}{d}\int_{-d/2}^{d/2}dx\,|\psi|^2.
\la{n}\eeq
Here $\psi$ is the wave function of the condensate, $m$ is the particle
mass, and the effective two-body interaction is given by $U_0 = 4 \pi
\hbar^2 a/m$, where $a$ is the scattering length for two-body collisions. The
condensate wave function satisfies the
nonlinear Schr\"odinger equation
\beq
-\frac{\hbar^2}{2m}
\frac{d^2\psi}{dx^2}+\left[V(x)+U_0|\psi|^2-\mu\right]\psi=0.
\la{Schr1}\eeq
The quantity $\mu$ is the chemical potential, and this depends both on the
mean density of particles, and on the quasimomentum. We shall look
for solutions where the mean density
is independent of position, and therefore these must satisfy the condition for
quasi-periodicity,
\beq
\psi(x+d)=\exp(i kd)\,\psi(x)
\la{quasiper}\eeq
for all $x$.
The quantity $\hbar k$ is the quasimomentum of the condensate. Our objective 
is to
find the energy per particle, the chemical potential, and the current density 
of nonlinear Bloch waves as a function of $k$ and of
$n$. The current density of the condensate
is
given by
\beq
j(k)=\frac{\hbar}{2mi}
\left(\psi^*\frac{d\psi}{dx}-\psi\frac{d\psi^*}{dx}\right).
\la{j}\eeq
It is independent of $x$ provided
$\psi$ satisfies \eq{Schr1}.
By direct calculation one can demonstrate that
\beq
j(k)=\frac{1}{\hbar}\frac{\partial {\cal E}(k, n)}{\partial k}.
\la{j2}\eeq
The quantity $j/n$ is the mean particle velocity.

Without loss of generality, we may choose $\psi(0)$ to be real. Since
the periodic potential is even in $x$, one infers from the
quasi-periodicity condition \ur{quasiper} that the real part of the
condensate wave function is an even function of $x$ and the imaginary part
an odd function.

The solution at the zone center ($k=0$) corresponds to a purely real and
periodic wave function, $\psi(x+d)=\psi(x)$.  Also at the zone boundary
($k=\pm\pi/d$) there is a purely real solution which is antiperiodic,
$\psi(x+d)=-\psi(x)$.  Since $\psi$ is real, these solutions have zero
current, $j=0$.  Other solutions are complex and the current is non-zero.
Since the potential in Eq.\ \ur{Schr1} is real, it follows that 
$\psi_k(x)=\psi^*_{-k}(x)$.   Naturally, all physical quantities
are periodic in $k$ with period $2\pi/d$.

The structure of the energy as a function of $k$ near the zone boundary
may be brought out by making a variational calculation
of the energy.  We use a trial function of the form
\beq
\psi(x)\!=\!\sqrt{n}\{\cos\alpha\, \exp{ikx}
+\sin\alpha \,\exp(i[k-2\pi/d]x)\},
\la{trial}\eeq
where $\alpha$ is a variational parameter.  This is similar to that used in 
the almost-free-electron  approximation for
the band structure of a single particle, and it also corresponds to the 
two-level 
model studied in Ref.\ \cite{WN1}.
When the trial function (\ref{trial}) is inserted into the energy functional 
\ur{calE}, it becomes 
\[
{\cal E}=n \left(\bar{\epsilon}
    +\frac{\Delta\epsilon}{2}\cos 2\alpha \right)+
\frac{nV_0}{2}\left(1-\half\sin 2\alpha\right)
\]
\beq
+\frac{n^2U_0}{2}\left(1+\half\sin^2 2\alpha\right),
\la{tot}\eeq
where the three terms are the kinetic, potential and
interaction energies, respectively, with $\bar{\epsilon}= (\epsilon_k + 
\epsilon_{k-2\pi/d})/2$, where $\epsilon_k=\hbar^2
k^2/2m$, and $\Delta\epsilon =\epsilon_k - \epsilon_{k-2\pi/d}
=(2\pi\hbar^2/md)
(k-\pi/d)$.
The optimal value of $\alpha$ is obtained by requiring \ur{tot}
to be stationary with respect to variations in $\alpha$,
and the chemical potential $\mu$ is then found by differentiating
the resulting energy with respect to the density, $\mu = \partial {\cal 
E}/\partial
n$.

At the zone boundary $k=\pm\pi/d$ the stationarity of \ur{tot}
yields two solutions: $\sin 2\alpha=V_0/2nU_0$ and $\cos 2\alpha=0$.
The first solution, which requires that the density exceed a critical value
$n_{\rm c}$ given by
\beq
n_{\rm c}=V_0/2U_0,
\label{nc}
\eeq
has an energy density
\beq
{\cal E}_1=n\left(\frac{\hbar^2\pi^2}{2md^2}
+\frac{V_0}{2}\right) +\frac{n^2U_0}{2}
-\frac{V_0^2}{16U_0},
\la{exactE}\eeq
a chemical potential
\beq
\mu_1=\mu_{{\rm c}}+(n-n_{\rm c})U_0,
\la{exactmu}
\eeq
where
\beq
 \mu_{\rm c}= V_0+\frac{\hbar^2\pi^2}{2md^2},
 \eeq
and a current density
\beq
j=\pm\frac{\hbar\pi}{md}\sqrt{n^2-n_{{\rm c}}^2}.
\la{exactj}
\eeq
The two possible signs for the current density 
reflect the
fact that if $\alpha$ is a solution to the stationarity condition, so is 
$\pi/2-\alpha$.

Equations (\ref{nc})--(\ref{exactj}) are in fact {\em exact}, the 
variational trial function
\ur{trial} being in this case identical with the analytical solution found in
Refs. \cite{BCDK,WDN}. At the critical density when the analytical solution
first
appears with $\alpha=\pi/4$, the current is zero since $\psi(x)$ is real. We
note that the critical chemical potential is the height of the barrier, plus
the kinetic energy of a free particle with a momentum equal to the
quasimomentum at the zone boundary.

For $n>n_{{\rm c}}$ the exact solution \ur{trial} with
$\sin 2\alpha = V_0/2nU_0 <1$ is complex and carries a
non-zero current. Remarkably, although this state has $k=\pm\pi/d$, it has 
neither the highest energy nor the highest chemical potential of all states in 
the
band: the {\em real}
antiperiodic wave function, also with $k=\pm\pi/d$, has higher
$\mu$, and it corresponds to the second solution of the stationarity
condition, $\cos 2\alpha =0$ or $\alpha=\pi/4$ (the other possibility,
$\alpha=3\pi/4$, belongs to the next, excited Bloch band and will not be
considered here). This solution exists at all densities and corresponds
to a chemical potential $\mu_2=\mu_{{\rm c}}+3(n-n_{{\rm c}})U_0/2$
which is higher than $\mu_1$ at $n>n_{{\rm c}}$.

For a general value of $k$, the $\Delta\epsilon$ term in Eq.\ (\ref{tot}) for
the
total energy tends to make the roots of the stationarity condition initially
at $\alpha = \pi/4$ and $\alpha = [\arcsin(V_0/2nU_0)]/2$ approach one
another, and the two roots merge and disappear for a sufficiently large value
of the magnitude of
$\Delta\epsilon$.  The energy per particle obtained using the trial function 
\ur{trial} is shown in Fig.\ 1,  where one sees the swallow-tail structure 
near the zone boundary.  Throughout, we shall measure energies in units of  
$\hbar \omega$, where $\omega =
(2 V_0/m)^{1/2}\pi/d$ is the angular frequency of small oscillations about a
minimum of the potential.

\begin{figure}[!htb]
\begin{center}
\resizebox *{6.5cm}{5cm}{\includegraphics{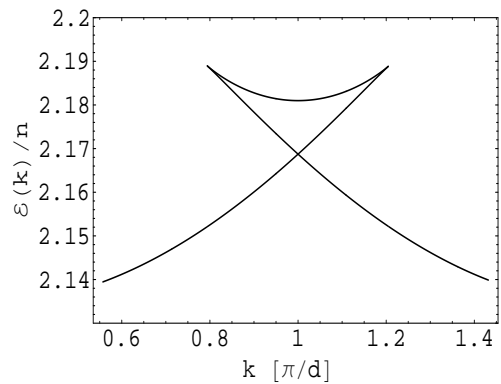}}
\end{center}
\caption{Swallow-tail structure of the energy per particle as a function of 
$k$ for $V_0=3\hbar \omega$ and  $n/n_{{\rm c}}$= 1.20.}
\end{figure}

A quantitative analysis shows that the trial form (\ref{trial}) is a 
good
approximation near the zone boundary provided $\hbar^2\pi^2/2md^2 \gg 
nU_0$.  The analytical calculations may be improved by including 
plane-wave components with wave numbers that differ from the ones in 
Eq.\ (\ref{trial}) by multiples of the reciprocal lattice vector $2\pi/d$.

We now demonstrate that the swallow-tail structure also arises in numerical 
solutions of the full Gross--Pitaevskii equation \ur{Schr1}.  At
condensate densities below the critical one, the wave function and dispersion
relation for nonlinear Bloch waves are qualitatively similar to those for a
single particle.  To illustrate this, we again consider the case where the 
potential barrier is
$V_0=3\hbar \omega$.

\begin{figure}[!htb]
\begin{center}
\resizebox *{6.5cm}{5cm}{\includegraphics{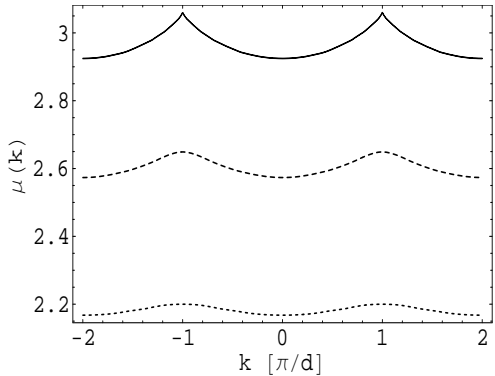}}
\end{center}
\caption{Chemical potential $\mu$  in units of $\hbar \omega$ as a function of 
quasimomentum for
densities $n/n_{{\rm c}}$= 0.57 (dotted), 0.78 (dashed), and 0.99 (full 
line).}
\end{figure}

\begin{figure}[!htb]
\begin{center}
\resizebox *{6.5cm}{5cm}{\includegraphics{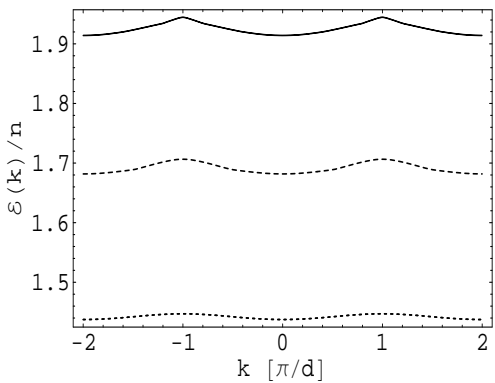}}
\end{center}
\caption{Energy per particle, 
${\cal E}/n$, in units of $\hbar \omega$ as a function of quasimomentum for
densities $n/n_{{\rm c}}$= 0.57 (dotted), 0.78 (dashed), and 0.99 (full 
line).}
\end{figure}

In Figs.\ 2 and 3 we plot the chemical potential $\mu(k,n)$ and the 
energy per particle for several
values of density $n<n_{{\rm c}}$. In all cases the chemical
potential lies below the top of the potential barriers. Therefore,
the energy bands are relatively narrow as the motion of the
condensate involves tunneling through the barriers. The difference between the 
chemical potential and the energy per particle reflects the nonlinear 
character of the interparticle interaction. 

\begin{figure}[!htb]
\begin{center}
\resizebox *{6.5cm}{5cm}{\includegraphics{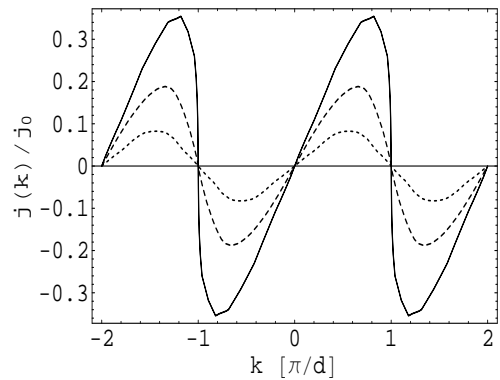}}
\end{center}
\caption{The current density $j$ as a function of quasimomentum at
densities $n/ n_{\rm c}$=0.57 (dotted), 0.78 (dashed), and 0.99 (full line).  
Here 
$j_0=\pi \hbar n/m d$ is the current density at  $k=\pi/d$ in the absence of a 
lattice.}
\end{figure}

In Fig.\ 4 we show the current density $j(k,n)$  of the nonlinear waves for 
the same number densities as
in Figs.\ 2 and 3. The current goes to zero both at the zone center and the 
zone 
boundary.
 At the critical density the real antiperiodic wave function corresponding
to the top of the band and having $k=\pm\pi/d$
coincides with the analytical solution,
\beq
\psi^{k=\pm\pi/d}_{{\rm c}}(x)=\sqrt{\frac{V_0}{U_0}}
\cos\frac{\pi x}{d}.
\la{exactcrit}\eeq
The current density $j(k,n_{{\rm c}})$ goes to zero at
$k=\pm\pi/d$ but has an infinite derivative at this point,
see Fig.\ 4. Consequently, $\mu(k,n_{{\rm c}})$ develops a singularity
at $k=\pm\pi/d$ (see Fig.\ 2), as does the energy density.

\begin{figure}[!htb]
\begin{center}
\resizebox *{6.5cm}{5cm}{\includegraphics{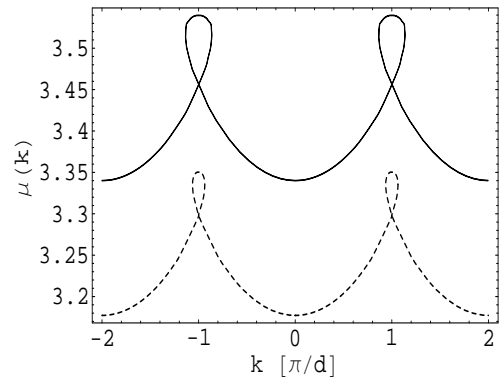}}
\end{center}
\caption{Chemical potential $\mu$ as a function of quasimomentum for
densities $n/n_{\rm c}=$ 1.14 (dashed) and 1.25 (full line).}
\end{figure}

For $n>n_{\rm c}$ the Bloch wave evolves from
a real periodic wave function at $k=0$ to the analytical
plane-wave solution \ur{trial} at $k=\pi/d$, which, however, is
now complex. At $k=\pi/d$ there is a second solution equal to the complex 
conjugate of the first one and, in addition, a third one
which is real, and which therefore carries no
current.  Also for $k$ close to $\pi/d$ one finds three solutions 
for a given $k$.  
  We plot the chemical
potential $\mu(k,n)$, the current density $j(k,n)$, and the
energy per particle ${\cal E}(k,n)/n$ in Figs.\ 5--7, respectively,
for two values of the density $n>n_{{\rm c}}$. We note that
$j(k,n)$ calculated from the wave function using Eq.\ (\ref{j}) is consistent 
with the relation (\ref{j2}). 

\begin{figure}[!htb]
\begin{center}
\resizebox *{6.5cm}{5cm}{\includegraphics{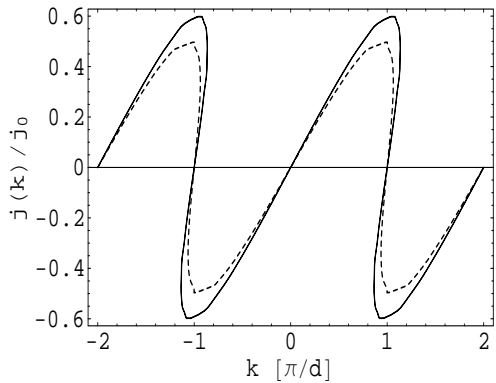}}
\end{center}
\caption{Current density $j$ as a function of quasimomentum for
densities, $n/ n_{\rm c}=$ 1.14 (dashed) and 1.25 (full line). The unit of 
current $j_0$ is 
the same as in Fig.\ 4.}
\end{figure}

\begin{figure}[!htb]
\begin{center}
\resizebox *{6.5cm}{5cm}{\includegraphics{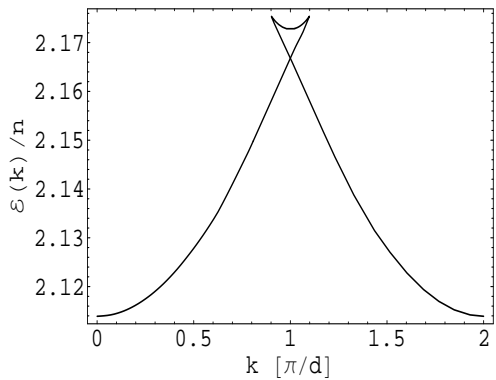}}
\end{center}
\caption{Energy per particle ${\cal E}/n$ of a Bloch wave as a function of
quasimomentum for a condensate density
$n/n_{\rm c}=1.20$.}
\end{figure}

The unusual structure of the energy bands has important implications for
experiment.  If a single particle in a periodic potential is subjected to a
small constant force, $k$ varies linearly in time, and the particle velocity
and position execute Bloch oscillations.  Such behavior has been observed in
optical lattices for thermal atoms above the Bose-Einstein transition
temperature \cite{salomon}.  Similar arguments apply to the motion of a
Bose--Einstein condensate, provided the energy of the condensate in the lowest
band is a continuous, single-valued function of $k$.  Bloch oscillations have
indeed been observed for condensates \cite{morsch}.  At densities greater than
$n_{\rm c}$, the situation is different because of the swallow-tail structure
of the energy band.  When a weak force is applied to a condensate initially at
rest, in the simple picture of the motion $k$ will increase and with time will
reach the value $\pi/d$.  With further increase of $k$, the system will
continue on the branch of the spectrum for $k>\pi/d$ which has the same slope
at $k=\pi/d$ as that of the initial branch.  With this choice of branch, the
wave function varies continuously as $k$ increases past $\pi/d$, whereas there
is a discontinuous change in the wave function of the
lowest energy state.  A dramatic manifestation of this is that the current
density in the lowest energy state changes sign at $k=\pi/d$.  With time, $k$
will eventually reach the tip of the swallow tail, and it then becomes
impossible to describe the condensate motion in terms of the usual adiabatic
picture, as was found in Ref.\ \cite{WN1} for a two-level model.  How to treat 
the dynamics under such conditions remains a challenging open problem.

 The experiments on Bloch oscillations in Ref.\ \cite{morsch} were carried out
in a regime in which the interaction energy was small compared with the depth
of the potential due to the optical lattice, and therefore one would not have
expected the energy band to have the swallow-tail structure.  It is important
to extend such experiments into the regime in which the strong nonlinear
effects predicted here should occur.  Theoretically, the stability of
a condensate moving in a periodic potential has been examined for relatively
low values of the two-body interaction \cite{WN2}, and such studies should be
extended to higher values of the interaction, where the swallow tail  
develops.

We are grateful to D.\ Aristov and S.\ J.\ Chang for very
useful discussions.

\end{document}